\documentclass{jps-cp}
\usepackage{txfonts} %Please comment out this line unless the txfonts package is availabe in your LaTeX system.
\usepackage{physics}
\usepackage{color}

\usepackage[normalem]{ulem}  % \sout{old text} for strikeout
\usepackage{color} % For blue in-text comments and additions

\renewcommand\sout{\bgroup \color{red} \ULdepth=-.5ex \ULset}
\title{
Threshold Cusp Structures in the Presence of Isospin Symmetry Breaking
}

\author{Katsuyoshi \textsc{Sone}$^{1}$ and Tetsuo \textsc{Hyodo}$^{2}$}

\inst{$^{1}$Department of Physics, Tokyo Metropolitan University, Hachioji 192-0397, Japan\\
$^{2}$Research Center for Nuclear Physics (RCNP), Ibaraki, Osaka 567-0047, Japan}

\email{$^{1}$sone-katsuyoshi@ed.tmu.ac.jp, $^{2}$hyodo@rcnp.osaka-u.ac.jp}

\recdate{February 21, 2019}

\abst{
We study the behavior of the cusp structures focusing on the isospin-breaking effects. The properties of the near-threshold exotic hadrons are encoded in the shapes of the cusp structures. In 
hadron scattering, 
it is often the case that
the thresholds of isospin partner channels are 
located within a narrow energy region. To analyze the scattering in such systems, it is therefore essential to study the cusp structures that emerge at two closely separated thresholds
with isospin symmetry breaking. 
In this study, we propose a practical 
representation
of the scattering amplitude and show that the two neighboring cusp structures are related through the isospin symmetry.
}

\kword{Threshold cusp, Low-energy scattering, Multichannel scattering}

\begin{document}
\maketitle

%%%%%%%%%%%%%%%%%%%%%%%%%%
\section{Introduction}
%%%%%%%%%%%%%%%%%%%%%%%%%%

Many exotic hadrons are known to appear in the vicinity of reaction thresholds. In the low-energy region
near the threshold, 
the scattering length 
is the key
physical quantity governing the scattering. 
Since the value of the scattering length is reflected in the shape of the threshold cusp structure~\cite{Esposito:2021vhu, LHCb:2021auc}, studying the behavior of threshold cusps is essential for understanding the properties of exotic hadrons~\cite{Baru:2004xg, Guo:2019twa, Sone:2024nfj, Sone:2025xuh}.

In some two-body hadronic systems, isospin symmetry causes two reaction thresholds to lie 
within a few MeV region. 
In such situations, the associated cusp structures emerge within a narrow energy window. In this work, we study how these cusp features behave in both isospin-symmetric and isospin-breaking situations, using the $\Lambda N$-$\Sigma N$ coupled-channels system as a typical example.

%%%%%%%%%%%%%%%%%%%%%%%%%%
\section{Formulation}
\label{eq: LN scattering}
%%%%%%%%%%%%%%%%%%%%%%%%%%

In this work, we consider $\Lambda N$-$\Sigma N$ coupled-channels scattering system, which includes three channels, $\Lambda p$, $\Sigma^{+}n$, and $\Sigma^{0}p$ with the total charge $Q=+1$. In the isospin basis, these channels correspond to $\Lambda N$-$\Sigma N$ with $I=1/2$ and $\Sigma N$ with $I=3/2$. The system involves two spin channels, singlet $J=0$ and triplet $J=1$. Because $\Lambda N$-$\Sigma N$ is the scattering between non-identical particles, each spin component contains both isospin $I=1/2$ and $I=3/2$ channels. In the following, we do not specify the spin component explicitly, and the results can be applied to either spin channel. In this system, the energy difference 
$\Delta_{\Sigma N}\sim 2$ MeV between the $\Sigma^{+}n$ and $\Sigma^{0}p$ thresholds is very small as a consequence of the isospin symmetry.

We focus on the threshold cusp structures in the $\Lambda p$ elastic scattering at the $\Sigma^{+}n$-$\Sigma^{0}p$ thresholds. First, we construct the scattering amplitude, which is described by a 3$\times$3 matrix in channel space. The optical theorem leads to the inverse of the scattering amplitude
\begin{align}
    f^{-1}
    &=
    \hat{K}^{-1} - i\hat{p},
    \label{eq: f inv}
\end{align}
where $\hat{K}$ is called the $K$-matrix and describes the interactions between the channels. In this analysis, we employ an isospin-symmetric $K$-matrix with real constants $C_{i}$:
\begin{align}
    \hat{K}
    &=
    \begin{pmatrix}
        C_1 & \sqrt{2}C_4 & C_4 \\
        \sqrt{2}C_4 & C_2 & \sqrt{2}(C_2 - C_3) \\
        C_4 & \sqrt{2}(C_2 - C_3) & C_3
    \end{pmatrix}.
    \label{eq: Kmatrix for lambdaN}
\end{align}
In a general three-channel scattering, the $K$-matrix contains six independent real parameters. However, the $K$-matrix~\eqref{eq: Kmatrix for lambdaN} includes only four parameters due to the isospin symmetry. The imaginary part of Eq.~\eqref{eq: f inv} is given solely by the momenta of each channel
\begin{align}
    \hat{p}
    &=
    \begin{pmatrix}
        p_{\Lambda p} & 0 & 0\\
        0 & p_{\Sigma^+ n}(E) & 0\\
        0 & 0 & p_{\Sigma^0p}(E)
    \end{pmatrix},
    \label{eq: img finv}
\end{align}
where $p_{\Lambda p},p_{\Sigma^+ n}(E)$ and $p_{\Sigma^0p}(E)$ represent the relative momenta of the $\Lambda p,\Sigma^+ n$, and $\Sigma^0p$ channels, respectively. We note that the momentum $p_{\Lambda p}$ is taken to be a constant, because the $\Lambda p$ threshold lies sufficiently far from the region near the $\Sigma N$ threshold which is of our interest. The momenta $p_{\Sigma^+ n}(E)$ and $p_{\Sigma^0 p}(E)$ depend on the energy $E$ and 
take different values at the same energy $E$ because of the small threshold
energy difference $\Delta_{\Sigma N}$. 
In the limit $\Delta_{\Sigma N}\to0$, where the $\Sigma^+ n$ and $\Sigma^0 p$ thresholds coincide, these two momenta become identical and the scattering amplitude in Eq.~\eqref{eq: f inv} is reduced to an isospin-symmetric form. Therefore, the matrix $\hat{p}$ provides the source of the 
isospin symmetry breaking in
the scattering amplitude arising from $\Delta_{\Sigma N}\neq0$~\cite{Sone:2025xuh}.

From Eqs.~\eqref{eq: f inv},~\eqref{eq: Kmatrix for lambdaN} and~\eqref{eq: img finv}, 
the $\Lambda p$ elastic scattering amplitude 
$f_{\rm el}(E)$
near the threshold of channel $i\ (i=\Sigma^+ n, \Sigma^0p)$ can be written as
\begin{align}
    f_{\rm el}(E)
    &=
    f^i_0\frac{1+ib_ip_i(E)}{1+ia_ip_i(E)}
    \quad
    (i=\Sigma^+ n, \Sigma^0p)
    \label{eq: fel}
\end{align}  
where $a_i$ represents the complex scattering length of channel $i$. We note that $b_{\Sigma^+ n}$ is real, whereas $b_{\Sigma^0 p}$ is in general complex. The coefficients $f^i_0$ in Eq.~\eqref{eq: fel} corresponds to the scattering amplitude at the threshold of channel $i$. Using Eq.~\eqref{eq: fel}, the $s$-wave cross section near the threshold of channel $i$ can be expanded in terms of the momentum $p_i(E)$ of channel $i$
\begin{align}
    \sigma_{\rm el}(E) &= 4\pi|f_0^i|^2 \left(1 + 2 \Im[a_i - b_i] p_i(E) + \mathcal{O}(p_i^2(E))\right), \quad (\text{above the threshold}), \label{eq: above ex csec}\\
    \sigma_{\rm el}(E) &= 4\pi|f_0^i|^2 \left(1 + 2 \Re[a_i - b_i] \kappa_i(E) + \mathcal{O}(\kappa_i^2(E))\right), \quad (\text{below the threshold}) .
    \label{eq: expansion of cses} 
\end{align}
The real momentum $\kappa_i(E)\equiv -ip_i(E)$ is introduced in Eq.~\eqref{eq: expansion of cses}, because $p_i(E)$ becomes pure imaginary below the threshold of channel $i$. From Eqs.~\eqref{eq: above ex csec} and~\eqref{eq: expansion of cses}, one can see that the slopes of the cross section above and below the threshold of channel $i$ are determined by $a_i$ and $b_i$. It should be noted that the slopes of the cross section at the threshold go to infinity when the energy $E$ is used as a variable. In this case, the shape of the cusp structure is determined only by the signs of the real and imaginary parts of $a_i-b_i$ ~\cite{Sone:2025xuh}. 

To clarify the contribution of the isospin-breaking effects $\Delta_{\Sigma N}$ in the slopes of the cross section, we expand $a_{\Sigma^+n}-b_{\Sigma^+n}$ and $a_{\Sigma^0p}-b_{\Sigma^0p}$ in terms of $\sqrt{2\mu_i\Delta_{\Sigma N}}$
\begin{align}
    a_{\Sigma^+n} - b_{\Sigma^+n} 
    &\simeq 2R + X\sqrt{2\mu_{\Sigma^0p}\Delta_{\Sigma N}}
    \quad 
    (\text{at the threshold of $\Sigma^+n$}),  
    \label{eq: slope of th 2} \\
    a_{\Sigma^0p} - b_{\Sigma^0p} 
    &\simeq R + W \sqrt{2\mu_{\Sigma^+n}\Delta_{\Sigma N}}
    \quad 
    (\text{at the threshold of $\Sigma^0p$}), 
    \label{eq: slope of th 3} \\
    R &\equiv -C_4^2/\{(1 - i C_1 p_{\Lambda p}) C_1\},
\end{align}
where $\mu_i\ (i=\Sigma^+n,\Sigma^0p)$ represents the reduced mass of channel $i$ and $X$ and $W$ are the complex constants which are determined by the $K$-matrix components and the constant momentum $p_{\Lambda p}$. The second terms in Eqs.~\eqref{eq: slope of th 2} and~\eqref{eq: slope of th 3} vanish in the isospin-symmetric limit $\Delta_{\Sigma N}\to0$ and therefore those terms represent the contributions of the isospin-breaking effects in the slopes of the cross section. The first terms in Eqs.~\eqref{eq: slope of th 2} and~\eqref{eq: slope of th 3}, which do not include $\Delta_{\Sigma N}$, are identical up to the factor $2$. This implies that the signs of the real and imaginary parts of $a_{\Sigma^+n} - b_{\Sigma^+n}$ and $a_{\Sigma^0p} - b_{\Sigma^0p}$ coincide with each other when the isospin-breaking effects are small: $2|R/X| \gg \sqrt{2\mu_{\Sigma^0p}\Delta_{\Sigma N}}$ and $|R/W| \gg \sqrt{2\mu_{\Sigma^+n}\Delta_{\Sigma N}}$. Consequently, it is expected that the identical types of the cusp structures appear at the thresholds of $\Sigma^+n$ and $\Sigma^0p$ for small isospin breaking effects. 

%We discuss the role of the factor $2$ multiplying $R$ in Eq.~\eqref{eq: slope of th 2}.
%When the energy $E$ is used as the variable, the slope of the cross section at the threshold diverges, and the factor $2$ determines the steepness of this divergence. Therefore, when the isospin-breaking effects are small, the cusp structure at the $\Sigma^+ n$ threshold is expected to be sharper than that at the $\Sigma^0 p$ threshold.

Next, we discuss the behavior of the threshold cusp structure in the isospin-symmetric limit $\Delta_{\Sigma N}\to0$. In this case, the thresholds of $\Sigma^+n$ and $\Sigma^0p$ are degenerate and only one cusp structure appears at the threshold of $\Sigma N$. In the same way as Eqs.~\eqref{eq: slope of th 2} and~\eqref{eq: slope of th 3}, the slopes of the cross section at the threshold of $\Sigma N$ is obtained as 
\begin{align}
    a_{\Sigma N} - b_{\Sigma N} &= 3R, \qquad (\Delta_{\Sigma N}\to 0),
    \label{eq: slope is}
\end{align}
where the scattering length $a_{\Sigma N}$ and complex constant $b_{\Sigma N}$ are defined as
\begin{equation}
\left\{
\begin{aligned}
    a_{\Sigma N} &= a_{\Sigma^+n}+a_{\Sigma^0p}, \\
    b_{\Sigma N} &= b_{\Sigma^+n}+b_{\Sigma^0p},
\end{aligned}
\qquad (\Delta_{\Sigma N}\to 0).
\right.
\label{eq: a b}
\end{equation}
Equations~\eqref{eq: slope is} and~\eqref{eq: a b} show that the slopes of the cross section in the isospin-symmetric limit is given by the sum of $a_{\Sigma^+n} - b_{\Sigma^+n}$ and $a_{\Sigma^0p} - b_{\Sigma^0p}$ with $\Delta_{\Sigma N}\to 0$. From this result, we therefore conclude that the two cusp structures at the thresholds of $\Sigma^+n$ and $\Sigma^0p$ merge into a single cusp in the $\Delta_{\Sigma N}\to 0$ limit.

%%%%%%%%%%%%%%%%%%%%%%%%%%
\section{Numerical results}\label{sec: num}
%%%%%%%%%%%%%%%%%%%%%%%%%%

In this section, we study the behavior of the threshold cusp structures numerically, focusing on the isospin-breaking effects. In this calculation, the $\Sigma N$ threshold energy is defined as the midpoint between the $\Sigma^{+}n$ and $\Sigma^{0}p$ thresholds, and this point is chosen as the origin of the energy. To discuss the behavior of the cusps, we focus on the $s$-wave cross section which is proportional to $|f_{\rm el}(E)|^2$ in Eq.~\eqref{eq: fel}. In this analysis, we use a dimensionless $s$-wave cross section 
\begin{align}
    \sigma_{\rm el}(E) 
    &= \frac{|f_{\rm el}(E)|^2}{|f_{\rm el}(E = 0;\, \Delta_{\Sigma N} \to 0)|^2}, 
    \label{eq: csec norm}
\end{align}
where $f_{\rm el}(E = 0;\, \Delta_{\Sigma N} \to 0)$ corresponds to the $\Lambda p$ elastic scattering amplitude in the isospin-symmetric limit. Accordingly, in the isospin-symmetric limit, the cross section $\sigma_{\rm el}(E;\Delta_{\Sigma N}\to 0)$ is normalized at the threshold of channel $\Sigma N$. The hadron masses used in this calculation are taken from Ref.~\cite{ParticleDataGroup:2024cfk}.

First, we consider a simplified model where the scattering amplitude is determined solely by the scattering length of $\Sigma N$ ($a_{\Sigma N}$). We impose the conditions $C_1 C_3 - C_4^2 = 0$ and $C_2 - 2 C_3 = 0$ on the parameters in Eq.~\eqref{eq: Kmatrix for lambdaN} to make the $K$-matrix~\eqref{eq: Kmatrix for lambdaN} separable. In this case, the scattering amplitude $f_{\rm el}(E)$ corresponds to the Flatt\'{e} amplitude which does not include background effects (see, e.g. Ref.~\cite{Sone:2024nfj}). Consequently, the behavior of resulting cross section is characterized only by the scattering length $a_{\Sigma N}$.

As a representative example, we set the scattering length $a_{\Sigma N}$ in Eq.~\eqref{eq: a b} as a typical hadronic scale
\begin{align}
    a_{\Sigma N} &= -1.0 - i\,0.8\quad {\rm fm}.
    \label{eq: a2s num}
\end{align}
With this scattering length, a quasivirtual pole~\cite{Nishibuchi:2023acl} is generated below the threshold of the $\Sigma N$ channel. For $a_{\Sigma N}$ in Eq.~\eqref{eq: a2s num}, the scattering lengths of $\Sigma^+n$ and $\Sigma^0p$ for the isospin broken case are calculated as
\begin{align}
    a_{\Sigma^+n} &= -0.65-i0.46 \quad {\rm fm},  \qquad
    a_{\Sigma^0p} = -0.26-i0.27\quad \rm fm. \label{eq: cp a3}
\end{align}
The normalized cross sections in Eq.~\eqref{eq: csec norm} are shown in Fig.~\ref{fig2}. The cross section in the isospin-symmetric limit ($\Delta_{\Sigma N}\to 0$) is shown by the dotted line and the solid line corresponds to the isospin broken one $\sigma_{\rm el}(E)$ with finite $\Delta_{\Sigma N}$.
 
 \begin{figure}[tbp]
    \centering
    \includegraphics[width = 8cm, clip]{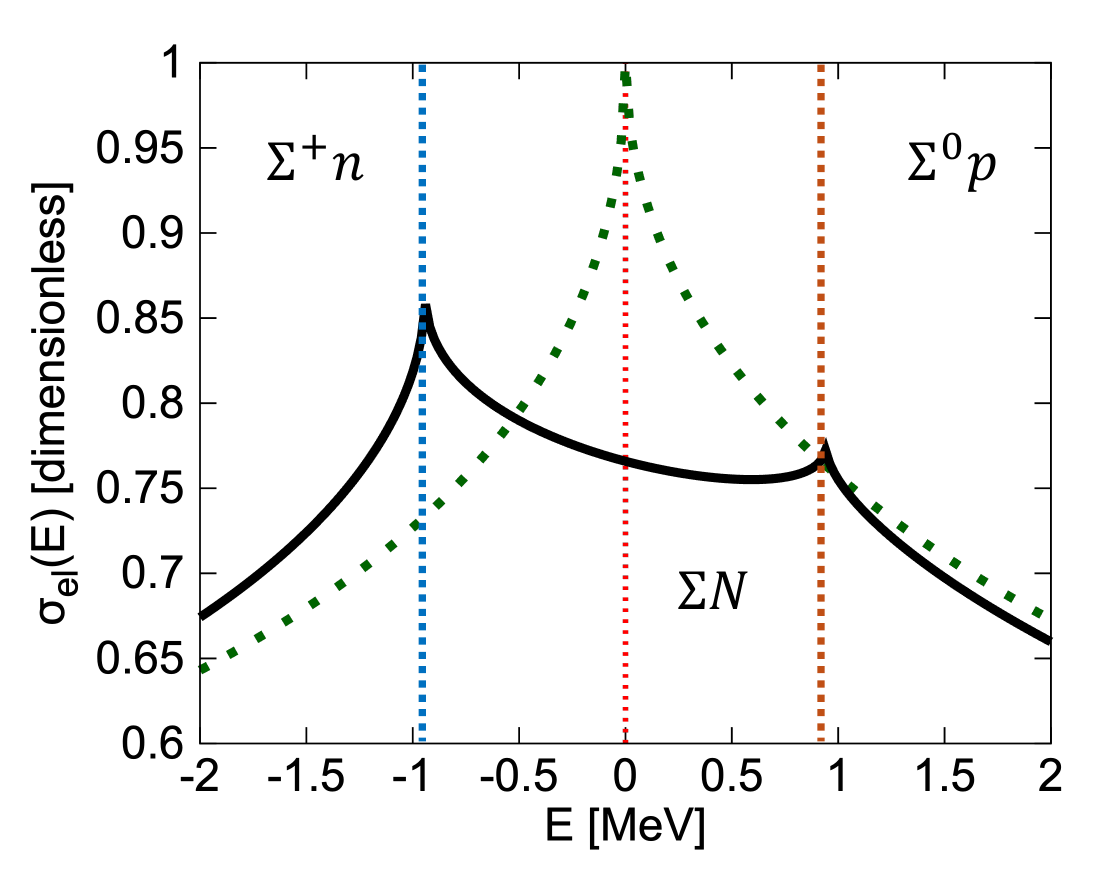}
    \caption{
        The normalized $\Lambda p$ elastic cross section $\sigma_{\rm el}(E)$ in Eq.~\eqref{eq: csec norm} with the scattering length $a_{\Sigma N}=-1.0-i0.8\ {\rm fm}$. The solid line corresponds to the isospin-broken cross section and the dotted line represents the isospin-symmetric one. The dotted vertical lines represent the thresholds of channel $\Sigma^+ n$, $\Sigma N$ and $\Sigma^0 p$.
        }
    \label{fig: Flatte}
\end{figure}

As seen in the solid line in Fig.~\ref{fig: Flatte}, the cross section $\sigma_{\rm el}(E)$ with finite $\Delta_{\Sigma N}$ exhibits similar upward cusp structures at the $\Sigma^+ n$ and $\Sigma^0 p$ thresholds. Moreover, we can see that the cusp at the $\Sigma^+n$ threshold is sharper than that at the $\Sigma^0p$ threshold. This behavior can be explained by the factor $2$ in Eq.~\eqref{eq: slope of th 2}. These results indicate that the isospin-breaking effects are small. In fact, the sum of $a_{\Sigma^+n}$ and $a_{\Sigma^0p}$ is nearly identical to the scattering length $a_{\Sigma N}$ in the isospin-symmetric limit in Eq.~\eqref{eq: a2s num}:
\begin{align}
    a_{\Sigma^+n}+a_{\Sigma^0p} = -0.91-i0.73\approx a_{\Sigma N}.
\end{align}
When the the isospin-breaking effect is small, the shape of the cusps is mainly determined by $R$ in Eqs.~\eqref{eq: slope of th 2}, \eqref{eq: slope of th 3}, and \eqref{eq: slope is}. Therefore, the dotted line in Fig.~\ref{fig: Flatte}, representing the isospin symmetric limit, also exhibits an upward cusp. 

%Consequently, we can see that the cusp structures at the $\Sigma^+n$ and $\Sigma^0p$ thresholds shows the identical types of the cusp structures for the small isospin-breaking effects. Moreover, as discussed in the previous section, the it is expected that the 

Next, we calculate cusp structures in the spin-triplet $\Lambda N$-$\Sigma N$ scattering with more realistic interaction. 
From now on, we deal with four parameters $C_{i}$ independently. The spin-triplet scattering lengths are taken from the chiral EFT analysis in Refs.~\cite{Haidenbauer:2019boi, Haidenbauer:2021smk, Haidenbauer:2023qhf}. In this case, the resulting cross section corresponds only to the spin-triplet component of the elastic $\Lambda p$ cross section, while the physical cross section generally includes both spin-triplet and spin-singlet contributions.

%Consequently, the $\Lambda p$ elastic cross section presented below also reflects only the contribution from the spin-triplet channel. 

In Ref.~\cite{Haidenbauer:2019boi}, the spin-triplet $\Sigma N$ scattering lengths are given for the NLO19 with cutoff $\Lambda=500$ MeV as
\begin{align}
    a_{1/2} = 0.95-i4.8 \quad {\rm fm}, \qquad
    a_{3/2} = -0.42\quad \rm fm, \label{eq: cp a3 H}
\end{align}
where $a_{1/2}$ is the scattering length for the total isospin $I=1/2$ and $a_{3/2}$ is that for $I=3/2$. Note that the state $\Sigma N$ with $I=3/2$ has no decay channels and therefore the imaginary part of $a_{3/2}$ is exactly zero in Eq.~\eqref{eq: cp a3 H}. From Eq.~\eqref{eq: cp a3 H}, three conditions are imposed on the four parameters in Eq.~\eqref{eq: Kmatrix for lambdaN}. To determine all parameters, we fix the parameter $C_1=2.0\ {\rm fm}$. The scattering lengths in the charge basis can be obtained from those in isospin basis by using the Clebsch-Gordan decomposition for $1/2\otimes3/2$
\begin{align}
    a_{\Sigma^+n} = \frac{1}{3}a_{3/2} + \frac{2}{3}a_{1/2}, \qquad 
    a_{\Sigma^0 p} = \frac{2}{3}a_{3/2} + \frac{1}{3}a_{1/2}.
\end{align}
The charge-basis scattering lengths corresponding to the chiral EFT results~\eqref{eq: cp a3 H}  are summarized in Table~\ref{tab: scattering_length_phys}. The corresponding $\Lambda p$ elastic cross sections are shown in Fig.~\ref{fig2}.

\begin{table}[t]
\centering
\caption{
Spin-triplet scattering lengths of $\Sigma N$ in the charge basis with Eq.~\eqref{eq: cp a3 H} and $C_1=2.0\ {\rm fm}$.
}
\label{tab: scattering_length_phys}
\begin{tabular}{c c c}
\hline\hline
Channel
 & Isospin symmetric (fm)
 & Isospin broken (fm) \\
\hline
$\Sigma^+ n$
 & $0.77-i3.2$
 & $-0.48-i3.0$ \\
$\Sigma^0 p$
 & $0.6-i1.6$
 & $0.52-i0.92$ 
 \\
\hline\hline
\end{tabular}
\end{table}

%The isospin-symmetric scattering lengths $a_{\Sigma^+n}$ and $a_{\Sigma^0p}$ in the charge basis for this parameter set are given by
%%\begin{align}
%%    a_{\Sigma^+n}^S&=0.77-i3.2\quad {\rm fm},\quad (\Delta_{\Sigma N}\to 0),\label{eq: a2 H} \\
%%    a_{\Sigma^0p}^S&=0.6-i1.6\quad {\rm fm}, \quad (\Delta_{\Sigma N}\to 0),\label{eq: a3 H}
%%\end{align}
%where the superscript S indicates the isospin-symmetric case. For $\Delta_{\Sigma N}\neq 0$, the scattering lengths in the isospin broken case are calculated as
%%\begin{align}
%%    a_{\Sigma^+n}&=-0.48-i3.0\quad {\rm fm}, \label{eq: a2 H bro} \\
%%    a_{\Sigma^0p}&=0.52-i0.92\quad {\rm fm}. \label{eq: a3 H bro}
%%\end{align}

\begin{figure}[tbp]
    \centering
    \includegraphics[width = 8cm, clip]{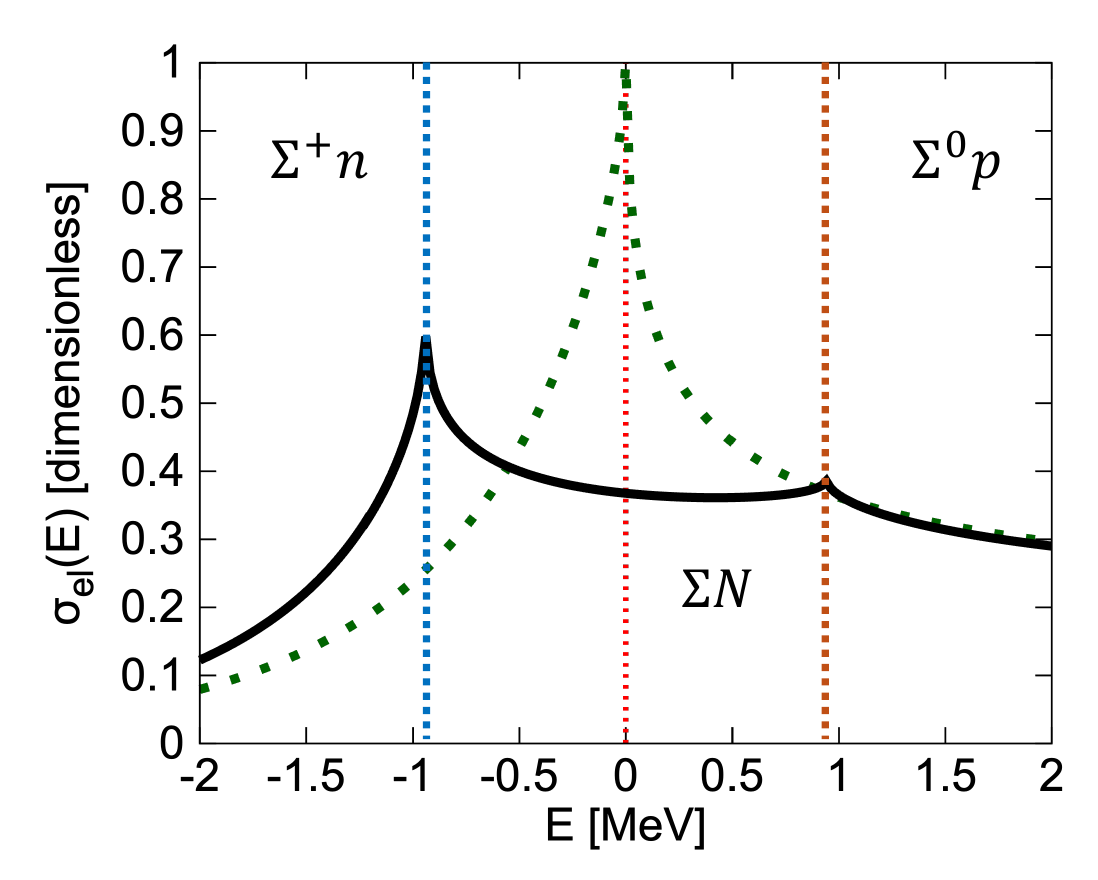}
    \caption{
        Same as Fig.~\ref{fig: Flatte} but the scattering lengths are fixed as Eq.~\eqref{eq: cp a3 H} with $C_1=2.0\ {\rm fm}$.
        }
    \label{fig2}
\end{figure}

From the solid line in Fig.~\ref{fig2}, we can see that the cross section with finite $\Delta_{\Sigma N}$ shows the upward cusp structures at both of the $\Sigma^+n$ and $\Sigma^0p$ thresholds. The dotted line in Fig.~\ref{fig2}, corresponding to the isospin-symmetric case, also shows the upward cusp at the $\Sigma N$ threshold. These qualitative features, namely the appearance of upward cusp structures at the thresholds, are similar to those observed in Fig.~\ref{fig: Flatte}. On the other hand, the difference in the sharpness of the cusp structures at the $\Sigma^+n$ and $\Sigma^0p$ thresholds is much more pronounced than that in Fig.~\ref{fig: Flatte}. Although a mild asymmetry is already present in Fig.~\ref{fig: Flatte}, it is strongly enhanced in Fig.~\ref{fig2}, where the $\Sigma^+n$ cusp becomes much more pronounced while the $\Sigma^0p$ cusp is strongly suppressed. Such a strong asymmetry in the sharpness of the $\Sigma^+ n$ and $\Sigma^0 p$ cusp structures is attributed to the effects of isospin breaking. In fact, the sum of the scattering lengths $a_{\Sigma^+n}+a_{\Sigma^0p}=0.04-i3.92\ {\rm fm}$ with finite $\Delta_{\Sigma N}$ is significantly different from $a_{\Sigma N}=1.4-i4.8$ fm for $\Delta_{\Sigma N}=0$; $a_{\Sigma N}$ is shifted by approximately $\sim1\ {\rm fm}$ due to the isospin-breaking effects. In other words, the isospin relation~\eqref{eq: a b} is broken largely.

%\begin{align}
%    a_{\Sigma^+n} + a_{\Sigma^0p} &= 3.8-i3.9\quad {\rm fm}, \label{eq: H sum}\\
%\end{align}
In this result, the isospin-breaking effects enhance the slopes of the cross section at the threshold of $\Sigma^+n$ while that at the threshold of $\Sigma^0p$ are suppressed:
\begin{align}
    a_{\Sigma^+n}-b_{\Sigma^+n}&=-2.7-i3.0 \quad {\rm fm}, \\
    a_{\Sigma^0p}-b_{\Sigma^0p}&=-0.54-i0.71 \quad {\rm fm}.
\end{align}
Again, we find large deviation from the isospin relation $a_{\Sigma^+n}-b_{\Sigma^+n}=2(a_{\Sigma^0p}-b_{\Sigma^0p})$. Consequently, the cusp structure at the $\Sigma^+n$ threshold is much stronger than that at the $\Sigma^0p$ threshold. In this way, the large isospin-breaking effects may change the behavior of the cusp structures while the two cusp structures are related in the isospin-symmetric limit.

%%%%%%%%%%%%%%%%%%%%%%%%%%
\section{Summary}
%%%%%%%%%%%%%%%%%%%%%%%%%%

In this contribution, we study the isospin-breaking effects on threshold cusp structures in coupled-channel hadronic scattering. First, in Sec.~\ref{eq: LN scattering}, we introduce a convenient representation of the scattering amplitude that enables a transparent analysis of the slopes of the cross section near thresholds. Applying this framework to the $\Lambda N$-$\Sigma N$ coupled-channel system, we analyze the cusp structures appearing in $\Lambda p$ elastic scattering in both the isospin-symmetric and isospin-broken cases. We show that, when the isospin-breaking effects are small, the cusp structures at the $\Sigma^+ n$ and $\Sigma^0 p$ thresholds exhibit similar shapes and merge into a single cusp in the isospin-symmetric limit. In Sec.~\ref{sec: num}, we show two numerical results: one corresponding to the case where the isospin-breaking effects are small, and the other to the case where they are sizable. For larger isospin-breaking, the sharpness of the two cusps differs significantly, reflecting the modifications of the slopes of the cross section induced by the isospin-breaking effects.

\section*{Acknowledgments}
This work has been supported in part by the Grants-in-Aid for Scientific Research from JSPS (Grants
No.~JP23H05439, % Kiban S (Hyodo)
No. JP22K03637, and % Kiban C (Hyodo)
 by JSPS KAKENHI Grant Number 25KJ1996, %Gakushin
and by MIYAKO-MIRAI Project of Tokyo Metropolitan University. % MIYAKO MIRAI PROJECT (Sone)

\bibliographystyle{JHEP}
\bibliography{refs.bib}

\end{document}